\documentclass[letterpaper, final, twocolumn, showpacs, pre,
tightenlines, lengthcheck, raggedbottom]{revtex4}

\usepackage{amsmath}
\usepackage{amsfonts}
\usepackage{amssymb}

\usepackage{mathptmx}

\usepackage{subfigure}

\usepackage{graphicx}

\usepackage{color}

\newcommand{\s}{\mathbf{s}}

\newcommand{\defeq}{:=}

\newcommand{\prob}[1]{\mathbb{P}_{\mathrm{eq}}\left( #1 \right)}

\newcommand{\size}[1]{\left| #1 \right|}

\newcommand{\mean}[1]{\left\langle#1\right\rangle}
\newcommand{\taurec}[1]{\tau^{\text{rec}}_{#1}}

 \newcommand{\E}{E}

\begin{document}

\title{Exact encounter times for many random walkers on regular and
complex networks}

\author{David P.~Sanders}
\email{dps@fciencias.unam.mx}
\homepage{http://sistemas.fciencias.unam.mx/~dsanders}
\affiliation{Departamento de F\'isica, Facultad de Ciencias, Universidad
Nacional Aut\'onoma de M\'exico,  Ciudad Universitaria,
04510 M\'exico D.F.,
 Mexico}
\affiliation{Centro de Ciencias de la Complejidad, Universidad
Nacional Aut\'onoma de M\'exico,  Ciudad Universitaria,
04510 M\'exico D.F.,
 Mexico}


\date{\today}

\begin{abstract}
 The exact mean time between encounters of a given particle 
in a system consisting of many
 particles undergoing random walks in discrete time is calculated, on both
regular and complex networks.  
 Analytical results are obtained both for independent walkers, where any number
of walkers can occupy the same site, and for walkers with an exclusion interaction,
when no site can
contain
more than one walker. These analytical results are then compared with
numerical simulations, showing very good agreement.
\end{abstract}

\maketitle

The last decade has seen an explosion of interest in the properties
and applications of complex networks with heterogeneous
 structure, due to their importance for modelling everything from 
social systems, such as the internet and networks of acquaintances, to 
biological ones, such as genetic regulatory networks 
\cite{BarabasiAlbertStatMechComplexNetworksRMP2002,
NewmanStructureFunctionComplexNetworksSIAM2003}.

After much initial work on the structure of these networks, attention has now
turned to 
dynamical processes which take place on them, with the aim of 
understanding the effect that different types of network structure have on the
dynamical properties of a system
\cite{NewmanStructureFunctionComplexNetworksSIAM2003,
BarratVespignaniDynProcesseComplexNetsBook2008}. As representative
examples in this direction, we mention studies on epidemics
\cite{BogunaAbsenceEpidemicthresholdScaleFreePRL2003}, the voter model
\cite{SoodRednerVoterModelHetGraphsPRL2005} and reaction-diffusion processes
\cite{PastorSatorrasColizzaRxnDiffnHetNetworksNatPhys2007} occurring on
complex networks. 

The properties of \emph{random walks} on networks have also attracted much
attention, both for single walkers 
\cite{NohRiegerRandomWalksComplexNetworksPRL2004,
BenAvrahamBolltDiffusionScaleFreeNetsNJP2005,
BenichouFirstPassageComplexScaleInvariantMediaNature2007}  and 
for multi-walker systems \cite{DeMouraFermiDiracStatsComplexNetworksPRE2005,
ArgyrakisBenAvrahamPriorityDiffnComplexNetworksPRE2008,
SandersLarraldeHowRareDiffusiveRareEventsEPL2008}.
These are perhaps the simplest systems involving motion of particles on
networks, and hence are of interest to understand the effect of the network
structure on diffusive properties such as mean transit time from one node to
another, and the mean time to return to a given node
\cite{BenAvrahamBolltDiffusionScaleFreeNetsNJP2005}.

These results have applications to \emph{individual-based} models,
in which ``agents'' (particles with internal states) diffuse  in
space until they \emph{encounter} each other, at which point they interact
following model-specific rules.
If the agents
neither die nor reproduce, then one of the key
quantities in the system is the time between these encounters, which we call
the \emph{encounter time}.

An example is the Bonabeau
model \cite{BonabeauPhaseDiagSelfOrganizingHierarchiesPhysA1995}, in which
agents represent animals which fight when they meet,
with the winner and loser gaining or losing social
status. 
This and similar models \cite{RednerDynamicsSocialDiversityJSM2005}
undergo a phase transition from a homogeneous,
non-differentiated society, to a society with two ``social classes'', one
successful and one unsuccessful
\cite{BonabeauPhaseDiagSelfOrganizingHierarchiesPhysA1995,
OkuboMFPhaseTransBonabeauPRE2007,
NaumisPhaseTransSocialHierarchiesAttractivePhysA2006}.
One of the key features of the analysis in this model is the timescale given by
the mean encounter time \cite{OkuboMFPhaseTransBonabeauPRE2007}.

Such models can also be studied on complex networks
\cite{GallosBonabeauScaleFreeIJMPC2005}.
Intuitively, for a complex network with
highly-connected hubs, all walkers have a tendency to migrate towards the hubs,
and thus they will encounter other walkers more frequently. The encounter
time provides a quantitative measure of this effect. 

Systems
of many particles undergoing random walks on complex networks are so
complicated that there are usually very few quantities which can be calculated
exactly.
Nonetheless, in this paper, it is shown that the mean encounter time of a given
walker in the system 
if often amenable to \emph{exact} calculation.

To calculate such mean encounter times, encounters are viewed in terms of
\emph{recurrences} (or returns) to a
set of encounter configurations, and the Kac recurrence theorem is applied. 
This theorem 
gives the exact recurrence time to a set in terms of its probability
in equilibrium, that is, the probability (frequency) of occupation of a set
after the system
has evolved for a long time and any transients have died away. For many random
walkers on complex
networks, even calculating such equilibrium probabilities already requires some
work \cite{ArgyrakisBenAvrahamPriorityDiffnComplexNetworksPRE2008}. The
calculation is also complicated by the necessity to carefully
define when encounters occur.  By carrying out these steps, we calculate
equilibrium
probabilities and mean encounter times for many random walkers with and
without
exclusion on regular and complex networks.

The paper is structured as follows. In sec.~\ref{sec:encounter-times}, 
some notation is introduced, and the main idea used in the paper is presented,
namely that 
encounter times may be expressed as recurrence times.
Sections~\ref{sec:networks-no-exclusion}--\ref{sec:complex-networks-exclusion}
treat in turn the cases of independent walkers on networks (regular or complex);
regular lattices with exclusion; and finally complex
networks with exclusion, which is
the least tractable case. Section~\ref{sec:conclusions} gives conclusions.

\section{Method and notation}
\label{sec:encounter-times}

We start by establishing some notation and describing the main technique to be
used throughout this paper.

\subsection{Encounter times and recurrence times}

We study a system of $N$ particles undergoing random walks on a
finite network. The network consists of $V$ nodes, with edges
joining them in a certain structure (see the next subsection).  We fix a
distinguished walker and assign to it the label ``$0$''.
The main question treated in this paper is how often this distinguished
walker ``interacts'' with other walkers, that is, what are its 
 \emph{encounter times} $\tau$. These are defined as the 
time intervals between the moments at which the distinguished particle meets
(encounters) other walkers, measured 
in terms of steps per particle (a ``sweep'').

This random variable $\tau$ has a certain distribution, which in
general 
is quite complicated. In this paper, we consider exclusively its mean
$\mean{\tau}$, which we call the 
\emph{mean encounter time} of the distinguished walker. 

The key idea to calculate mean encounter times
is the following: encounters of a given walker
correspond exactly to
\emph{returns}, or recurrences, to a particular set, namely the set $\E$
of configurations of the $N$ walkers for which an encounter of walker $0$
occurs.   A similar method was recently applied in a related context 
in ref.~\cite{SandersLarraldeHowRareDiffusiveRareEventsEPL2008}.

The spatial configuration of walkers is given by $\s \defeq (s_1,\ldots, s_N)$
specifying the location (site) $s_j$ of each walker $j$.
To describe encounters, however, the spatial locations are not 
sufficient -- we must also specify which walkers
are chosen to interact at a given time.  The rule for doing so is part of the
definition of a given model. 
We call the combination of the spatial and interaction information an 
\emph{extended} configuration.

The mean encounter time $\mean{\tau}$ of walker $0$ is then given exactly by
the \emph{mean
recurrence
time}  $\mean{\taurec{E}}$ to the set $\E$ of extended configurations 
corresponding to that walker's encounters: $\mean{\tau} = \mean{\taurec{E}}$.
We can thus make use of the \emph{Kac recurrence theorem}
\cite{KacRecurrenceThmBullAMS1947, KacProbabilityPhysicalSciencesBook,
CondaminRandomWalksFirstPassageConfinedPRE2007,
AldousFillRandomWalksGraphsBook}, which
gives an \emph{exact} result for the 
mean recurrence time $\mean{\taurec{A}}$ to a set $A$ in an ergodic,
discrete-time system, namely \begin{equation}
\mean{\taurec{A}} = \frac{1}{\prob{A}},
\label{eq:kac-original}
\end{equation}
where $\prob{A}$ is the probability that the system is in 
$A$ in equilibrium.

Calculating the mean encounter time
 thus reduces to the calculation of the equilibrium probability $\prob{\E}$
of the encounter set.
Note that higher moments and
other features of the complete probability distribution of recurrence
times are in general much harder to calculate
\cite{CondaminRandomWalksFirstPassageConfinedPRE2007,
AldousFillRandomWalksGraphsBook},
and will not be addressed here.

A special case is that of 
systems in which
the transition probabilities 
$P_{\mu \to \nu}$ 
from one configuration, $\mu$, to another, $\nu$,
 are symmetric, satisfying
$P_{\mu \to \nu} = P_{\nu \to \mu}$. The condition of detailed balance, which
holds throughout the paper, states that the flux of probability from $\mu$ to
$\nu$ in equilibrium is equal to that in the reverse direction:
\begin{equation}
 p_\mu P_{\mu \to \nu} = p_\nu P_{\nu \to \mu}.
\end{equation}
Thus, systems with symmetric transition probabilities have equal
equilibrium
probability for all (accessible) configurations.
In this case, the Kac result thus reduces to
$\mean{\taurec{A}} = 
\size{\Omega} / \size{A}$,
where $\Omega$ is the set of all microscopic configurations of the system, and
$\size{\cdot}$ denotes the cardinality (number of elements) of its argument.

Note that the encounter time as we have defined it above is
a single-particle quantity. Since all walkers are equivalent, it may also
be calculated by multiplying by $N$ the mean interval between encounters
involving \emph{any} of the walkers in the system.

\subsection{Notation for network structure}
Throughout the paper, the fixed number of walkers is denoted by $N$, 
the finite number of nodes in the undirected network by $V$, and the mean
density of walkers per node by $\rho \defeq N/V$. 
General references for network structure
include refs.~\cite{BarabasiAlbertStatMechComplexNetworksRMP2002,
NewmanStructureFunctionComplexNetworksSIAM2003}.

The sites of the network are labelled by $i$, and 
the \emph{degree} (number of incident edges) of the site $i$ is denoted by
$k_i$. The total
number of
sites in
the network with degree $k$ is $n_k \defeq \sum_i \delta_{k_i, k}$,
and the
\emph{degree distribution} is then $P(k) := n_k / \sum_k n_k $, which is the
probability that a randomly-chosen site has degree $k$. 

Finally, $K \defeq \sum_i k_i$  denotes the
total number of
outward edges in the network, which is twice the total number of undirected
edges, since each is counted twice, and  $\mean{k} \defeq K / V$ is
the mean number of incident edges per site, which satisfies $K = V \mean{k}$
and $N /
K
= \rho / \mean{k}$.

\section{Independent walkers on regular and complex networks}
\label{sec:networks-no-exclusion}

 The conceputally simplest case is that of 
many \emph{independent} random walkers on regular or complex networks,
with dynamics given as follows.
At each time step, a
single one of the $N$ walkers is selected at random (uniformly). If this
walker is
at site $i$, then it chooses
one of its $k_i$ neighbouring sites randomly (uniformly), and jumps to it.

Under these dynamics, each walker is independent, and thus the known results
for 
single walkers performing random walks on complex networks 
 can be applied: each
walker spends a proportion of time $k_i/K$ at node $i$
\cite{NohRiegerRandomWalksComplexNetworksPRL2004,
AldousFillRandomWalksGraphsBook} (i.e., a time
proportional
to the degree of the node). Recall that  $K = \sum_i
k_i$ is the total degree sum.

\subsection{Equilibrium distribution}

First let us consider the exact equilibrium distribution of the occupation
number at
a site,
i.e., the probability $q_i(m)$ 
of having $m$ particles at a given site $i$ of degree $k_i$.
Since the walkers are independent, and visit site $i$ with probability $p_i =
k_i / K$, the probability $q_i(m)$ that site $i$ has occupation number $m$ is
given by 
the following binomial
distribution:
\begin{equation}
q_i(m) 
= \binom{N}{m}
\left(\frac{k_i}{K}\right)^m \left(1 - \frac{k_i}{K}\right)^{N-m}.
\end{equation}
The mean occupation number $\mean{n_i}$ of site $i$ is then given by the mean of
distribution:
\begin{equation}
 \mean{n_i} = N p_i = \frac{N k_i}{K} 
= \mu  k_i,
\end{equation}
where we have defined $\mu \defeq \rho / \mean{k}$.
Thus $\mean{n_i}$ is proportional to $k_i$.

In the limit of infinite system size, $V \to \infty$ with $N \to \infty$ but
$\rho \defeq N/V$ fixed, we obtain asymptotically a Poisson distribution:
\begin{align}
q_i(m) \sim 
\frac{1}{m!}(\mu k_i)^m \exp(- \mu
k_i),
\end{align}
which  is the approximate result obtained in  
\cite{ArgyrakisBenAvrahamPriorityDiffnComplexNetworksPRE2008}.

\subsection{Mean encounter time}

The mean encounter time of a distinguished walker,
labelled by $0$, 
is calculated using the equilibrium probability $\prob{E}$ of the walker's
encounter set.  When there is no
exclusion, and
several particles may occupy the same site, there are several possible
definitions of when encounters occur; here, the following one is chosen.
If the walker which moves lands at an unoccupied site, then no encounter
occurs.  If, however, the walker lands at a site containing $m$ other
particles, then the moving walker chooses exactly \emph{one} of those $m$
particles and interacts with it, i.e., each particle in this pair of
walkers undergoes an encounter, but no other particle does so. This allows at
most a single
encounter at each time step, and forces the encounter to be a result of
movement.

Consider the set of encounter configurations defined \emph{after} the walker
has moved, and which contain information about which walker moved and which
other walker (if any) is involved in the interaction, apart from the spatial
positions of each walker. These encounter 
configurations are of two types:
(i) those 
in which the distinguished walker was chosen to move, and it moved to a
site which already contained at least one walker; and (ii) those in which a
walker other than the distinguished one moved, this walker landed on the site
which
contains the distinguished walker, and furthermore the distinguished walker was
chosen as the interaction partner of the moving walker.

In each of these two cases, the probability that the distinguished walker
interacts is $1/N$ times the probability that there is at least one other
walker on the same site as the distinguished walker after the jump. This is
clear in case (i). For case (ii), suppose
that there are $m_i$ walkers at the site, one of which is the distinguished
walker $0$, but that $0$ was not the moving walker. Then one of the other $m_i
-1$ particles is the one that was chosen to move, with total probability $(m_i
-1)/ N$, and after arriving at the new site an interaction was selected with
the distinguished walkers $0$, with probability $1/(m_i - 1)$. The total
probability is the product of these, so we regain the same expression.

It remains to calculate the probability that there is at least one
other walker on the same site as the distinguished walker 
The equilibrium probability that the distinguished particle is on a given site
$i$ with a total of $m$ walkers at that site (including the distinguished
one) is given by
\begin{equation}
\frac{k_i}{K} \binom{N-1}{m-1} \left(\frac{k_i}{K}\right)^{m-1}
\left(1 - \frac{k_i}{K}\right)^{(N-1)-(m-1)}.
\end{equation}
The term $k_i / K$ denotes the probability that the distinguished
walker is at site $i$, the second term represents the fact that there are $m-1$
other walkers at
the same site, and the last term represents the fact that the remaining $N-m$
walkers are at some other site.
The probability that the distinguished particle interacts is given by the
previous expression multiplied by $2/N$, provided $m>1$.

The encounter probability if the distinguished particle is at site
$i$ can thus be calculated as $2/N$ times the probability that the distinguished
particle is not alone at that site:
\begin{equation}
 \frac{2k_i}{NK} \left[ 1 - \left(1 - \frac{k_i}{K} \right)^{N-1} \right].
\end{equation}
The total encounter probability is then given by a sum over
all sites $i$:
\begin{equation}
 \prob{\E}= \frac{2V}{NK} \sum_i k_i  \left[ 1 - \left(1 -
\frac{k_i}{K} \right)^{N-1} \right],
\end{equation}
finally giving the exact result for the mean encounter time per
particle:
\begin{equation}
\mean{\tau} = \frac{1}{N \prob{\E}} = 
\frac{\mean{k}}{2
\mean{k \left[ 1 - \left(1 -
\frac{k}{K}
\right)^{N-1} \right]}_k}.
\end{equation}
Here, $\mean{\cdot}_k \defeq \sum_k [P(k) \cdot]$ 
denotes the mean of its argument over
the degree distribution. We have divided
by $N$ to give the physical time, such that each particle moves on average once
per time step.
Asymptotically for $N\to\infty$ with $\mu$ fixed, we obtain
\begin{equation}
\mean{\tau} \sim \frac{\mean{k}}{2 \mean{k \left[ 1 -
\exp\left(-\mu k \right) \right]}_k}.
\end{equation}

For regular networks with constant coordination number $z$, the degree
distribution is $P(k) = \delta(k-z)$.  For such networks, we thus
obtain
\begin{equation}
\mean{\tau} = \frac{1}{2} \left[ 1 - \left( 1 - \frac{\rho}{N}
\right)^{N-1}
 \right]^{-1} \sim \frac{1}{2 \left[ 1 - \exp(-\rho) \right] },
 \label{eq:independent-regular-network}
\end{equation}
which is again \emph{independent} of the coordination number $z$.

\section{Regular networks with exclusion}
\label{sec:reg-lattices-exclusion}

We now turn to walkers interacting via an exclusion interaction,
so that each site can be 
occupied by at most one walker
\cite{DeMouraFermiDiracStatsComplexNetworksPRE2005}. In this section we
consider the dynamics on a \emph{regular} network, i.e., one in which each site
has the same degree (number of neighbours), denoted by $z$. The
best-known subclass of such networks  consists of regular lattices;
other regular networks include 
small-world networks with a constant number of links per node.

The dynamics are as follows.
Initially, the walkers are distributed uniformly on the lattice, but
such that there is at most one walker at each site.  The dynamics
maintain this restriction, as for example in the
Bonabeau model discussed in the introduction
\cite{BonabeauPhaseDiagSelfOrganizingHierarchiesPhysA1995}, and are defined as
follows.
At each time step, a walker is picked at random.  This walker attempts to move
to one of its $z$ neighbouring sites, each with equal probability.  
If the trial site is empty, then the walker moves to the new site.  If the trial
site is
occupied by another walker, however, then the walkers interact, an
encounter occurs, and the invading particle remains where it
is, without moving.

Since the transition probabilities between two configurations are symmetric,
all configurations $\s$ have the same equilibrium probability, $\prob{\s} = 1 /
\size{\Omega} = (V-N)! / V!$.
More generally, we could allow the two interacting walkers to exchange
positions with (fixed) probability $p_{\text{exch}}$. 
The equilibrium probability $\prob{\s}$, and hence also the mean encounter time
$\mean{\tau}$, are unaffected by this.

\subsection{Mean encounter time}
\label{sec:reg-lattice-exclusion-mean-encounter}

\subsubsection{Mean-field argument for encounter time}

The following simple mean-field argument has been used to
estimate the mean encounter time in this system, in
refs.~\cite{BonabeauPhaseDiagSelfOrganizingHierarchiesPhysA1995,
OkuboMFPhaseTransBonabeauPRE2007}.

At each time step, a single walker moves, each with probability $1/N$, so that
the distinguished walker moves on average once every $N$ steps.
Suppose that the distinguished walker $0$ does move. The probability of
an encounter is the probability that the site it jumps to (one of its $z$
neighbours) is occupied, which is $\rho$, assuming that all walkers are
distributed uniformly on the lattice (the mean-field assumption).
Similarly, another walker can attempt to move to the site where walker $0$ is
sitting. The total probability of the distinguished walker interacting is thus
$2
\rho$, giving an estimate
\begin{equation}
\mean{\tau} \simeq 1/(2 \rho) = V/(2N)
\end{equation}
for the mean encounter time per particle.
We can refine this calculation by taking $(N-1)/(V-1)$, rather than $\rho$,  as 
the probability that the site
jumped to is occupied, by conditioning on
the fact that the departure site is occupied. This gives
$\mean{\tau} \simeq (V-1)/2(N-1)$.
Note that this mean-field calculation is also appropriate for the dynamics
without exclusion studied in the last section, in the case of a regular
network. Indeed, expanding \eqref{eq:independent-regular-network} for small
$\rho$ gives back this mean-field result.

The above argument gives an (uncontrolled) approximation of the
mean encounter time
$\mean{\tau}$ on a lattice. It is not clear, however, how good an
approximation it is.
We now show that $\mean{\tau}$ can in fact be calculated 
\emph{exactly} 
using the approach of section~\ref{sec:encounter-times}.
 The result of the (refined) mean-field calculation 
turns out to be exactly correct, suggesting that when we average over all
possible configurations of the particles, 
space ``no longer matters''.

\subsubsection{Exact calculation of encounter time}
For a regular network with exclusion, 
all microscopic configurations are equally likely,
as shown above, so that the Kac recurrence theorem gives
 $\mean{\tau} = \size{\Omega}/\size{\E},$
where $\E$ again denotes the encounter set of extended
configurations for which the distinguished walker $0$ undergoes an encounter.

To calculate the mean encounter time $\mean{\tau}$ of a distinguished walker,
we must first explicitly define
the set $E$ of encounter 
configurations.
It is not initially clear how to do this, since
two walkers can never occupy the same site. 

In fact, an encounter occurs exactly when the walker
which is chosen to
move does so towards an occupied site.  To indicate this direction of motion, we
augment the positional configuration  of
the particles $\s$ (before the move) with an arrow which sits on top of the
moving
walker and points in its chosen direction of motion -- one of $z$ possible
directions.
The extended configurations thus take the form $(\s; w, d)$, where $w \in \{1,
\ldots,N\}$ is the
label of the moving walker and $d \in \{1,\ldots,z\}$ is its chosen direction.

The set $\Omega$ of all extended configurations is thus given by assigning to
each of the $N$ walkers a distinct
site out of the $V$ possible sites, 
choosing  one of the $N$ occupied sites  as the moving walker, and then
choosing one of $z$ possible directions of motion.
The total number of extended configurations is thus
\begin{equation}
 \size{\Omega} = \frac{V!}{(V-N)!} Nz.
\end{equation}

An encounter involves 
the distinguished walker $0$ if either (i) walker $0$ is chosen to
move, and it attempts to move towards an occupied site; or
(ii) walker $0$ occupies the site towards which another walker attempts to move.
The first requirement for the set $\E$ is thus that the distinguished walker
has at least one neighbouring site occupied.

We split the set $\E$ of encounter configurations of the 
distinguished walker into disjoint sets $\E_p$, in which this 
walker has exactly $p$ out of its $z$ neighbouring sites occupied
(with
$1 \le p \le z$) and it actually does encounter a neighbouring
walker, i.e., walker $0$ moves towards an \emph{occupied} neighbour, or one of
the particles in the neighbouring sites jump towards walker $0$. Note that the
sets $\E_p$ do not fill up
the whole of
$\Omega$, due
to these
jumping conditions.

 The mean encounter time per particle of the distinguished walker $0$ is
thus given by
\begin{equation}
\mean{\tau} = \frac{\size{\Omega}}{N \size{\E}}
= \frac{\size{\Omega}}{N \sum_{p=1}^z \size{\E_p}}.
\end{equation}
The calculation of the sizes $\size{E_p}$ of the sets $E_p$
proceeds via the following 
combinatorial arguments.

First consider $\E_1$, the configurations in which the distinguished
walker $0$ has a single occupied site and does encounter its single
neighbouring particle when one of them moves.
Walker $0$ can
be placed in any of the $V$ sites; the single neighbour can then be chosen from
the
other $(N-1)$ walkers, and placed in any of the $z$ neighbouring
sites. 
The $(N-2)$ remaining walkers can be placed in any of the remaining
$(V-(z+1))$
sites.
Finally, only two configurations of the arrow are allowed: one which points
from the distinguished walker to its single occupied neighbour, and another
pointing from the neighbour to the distinguished walker.
Thus
\begin{equation}
 \size{\E_1} = 2zV  (N-1)   \frac{[V-(z+1)]!}{[V-(z+1)-(N-2)]!}.
\end{equation}

Similarly, when the site of the walker $0$ has $p$ occupied neighbours we
obtain
\begin{equation}
 \size{\E_p} = 2pV \binom{z}{p}  \frac{(N-1)!}{(N-1-p)!}
\frac{[V-(z+1)]!}{[(V-N)
- (z-p)]!}.
\end{equation}
Here, the binomial coefficient $\binom{z}{p}$ counts the number of ways of
choosing the $p$ neighbouring sites out of $z$ to be occupied, the number of
permutations $\frac{(N-1)!}{(N-1-p)!}$ gives the number of ways of placing $p$
of the remaining $(N-1)$ walkers in those neighbouring sites, and the arrow can
be in any of $2p$ configurations.
(These results remain valid for $N$ close to $0$ or close to $V$ if we define
the permutations to be $0$ when the number to choose is greater than the number
available.)

The expression for $\size{\E_p}$ may be rewritten as
\begin{align}
 \size{\E_p}
= 2Vz
 (N-1)! \binom{z-1}{p-1} \binom{V-(z+1)}{N-(p+1)}.
\end{align}
Thus, setting $u \defeq p-1$, we have
\begin{align}
 \sum_{p=1}^z \size{\E_p} &= 2Vz (N-1)! \sum_{u=0}^{z-1} \binom{z-1}{u}
\binom{V-(z+1)}{(N-2) -u} 
\label{eq:sum-first}\\
&= 2Vz (N-1)!  \binom{V-2}{N-2}.
\label{eq:sum-second}
\end{align}
The equality in \eqref{eq:sum-second} comes from the interpretation of the
sum in \eqref{eq:sum-first} as the number of ways of choosing $(N-2)$ boxes from
a total
of $(V-2)$, split into a choice of $u$ from the first $z-1$ boxes, and the
remaining $(N-2)-u$ from the $V-(z+1)$ remaining boxes.

Finally, we obtain the \emph{exact} result
\begin{equation}
\mean{\tau} = \frac{V-1}{2(N-1)}.
  \label{eq:exact-encounter-time-reg-exclusion}
\end{equation}
Note that this is independent of the coordination number $z$, and hence is
valid for \emph{any} regular network.

One might think that the independence under the dynamics of the Kac
result would immediately show that the spatial and
mean-field results are the same. 
In fact, however, the
above argument shows that 
the sets of \emph{extended} encounter configurations differ in each case,
and so the argument is more involved -- despite the simplicity of the final
result,
there does not appear to be a simpler derivation.

\subsection{Including a probability of interaction}

Within this same framework, we can allow for the
possibility that actual encounters occur only
a certain fraction $p$
of the
time, even if particles meet. 
This could model a repulsion between agents, so that there is an unwillingess to
interact, or a territory that is large enough so that two animals in the same
coarse-grained cell move past each
other without seeing each other.

To calculate the mean encounter time in this case, the configurations may be
extended further, taking the form $(\s; w, d, b)$, where the $b$ are
independent
Boolean variables $b \in \{0,1\}$ which indicate whether or not an encounter
occurs.  

Denoting the new encounter set by $E' \defeq E \times \{1\}$, where $\{1\}$ 
denotes when the Boolean
variables are true, the result is $\prob{E'} = p \prob{E}$, and
hence $\mean{\tau'} = \frac{1}{p} \mean{\tau}$, so that the effect of including
the probability $p$
is an extra factor $\frac{1}{p}$ in the expression for the mean encounter time, 
as is intuitively expected.

\section{Complex networks with exclusion}
\label{sec:complex-networks-exclusion}

In this section, we extend the results for many random walkers with an
exclusion interaction to the case of complex networks, with heterogeneous
degree distribution. 
The dynamics is as follows. At each step, a walker is
selected uniformly. If the walker is at site $i$, then it attempts to jump
to one of its $k_i$ neighbouring sites, with equal
probabilities $1/k_i$. If the trial site is unoccupied, then the jump is
allowed,
and the particle is moved; if the trial site is occupied, then the jump is
rejected,
 and the particle remains where it was.

This case was previously studied in
ref.~\cite{DeMouraFermiDiracStatsComplexNetworksPRE2005}
by viewing the system as fermions and relating the equilibrium distribution of
occupation numbers to  the Fermi--Dirac distribution. 
Here we reconsider these results using a more intuitive method from
ref.~\cite{ArgyrakisBenAvrahamPriorityDiffnComplexNetworksPRE2008}.

The combination of a heterogeneous network and the exclusion
interaction
makes the calculation of even the equilibrium occupation number distribution
highly non-trivial; indeed, it does not appear to be possible to obtain simple,
exact
results for this quantity in general
\cite{DeMouraFermiDiracStatsComplexNetworksPRE2005}. In the next section we show
that exact
results can
be obtained in the case of simple, structured networks. In the following section
we then
give an approximate argument valid for large networks.

\subsection{Small, structured networks}

For complex networks with some structure or which are small enough, it
is possible to obtain exact results for the complete equilibrium distribution
$\prob{\s}$ for each microscopic configuration $\s$, and from there obtain
coarse-grained quantities such as the mean occupation number of a given node,
by
explicit calculation. Here we give a
simple example, which illustrates the general method.

We consider a star-shaped network, representing
a single hub in a
complex network. The network consists of a central site $0$ with $L$ links
to sites
$1,
\ldots, L$, each of which has only a single link back to the hub. 
We consider two walkers moving with exclusion on this network, so that the
possible 
configurations $\s$ are of the form $\s = (s_1, s_2)$, where $s_j$ is the site
occupied by particle
$j$, although the results are easily extended to more particles. There are two
types of configuration: those with a
particle at the hub, of the form $\s =
(0, i)$ or $(i,0)$, of which there are $2L$; and those with no particle at the
hub, of the form $\s = (i,j)$, with $i,j \ge 1$ and $i \neq j$, of which there
are $L(L-1)$.

\subsubsection{Equilibrium probability}
Let $p \defeq p_{(i,j)}$ be the equilibrium probability to be in configuration
$(i,j)$, with $i \neq 0$ and $j \neq 0$, i.e., with no particles at the hub). 
All probabilities are symmetric in
the two arguments.
The transition probabilities are given by
\begin{equation}
 P\left[ (0,i) \to (j,i) \right] = \frac{1}{2L}; \qquad 
 P\left[ (j,i) \to (0,i) \right] = \frac{1}{2}.
\end{equation}
The second equation follows from the fact that there is a probability $1/2$ to
move each particle from a configuration $(j,i)$, to arrive at the
configuration
$(0,i)$ or $(j,0)$.  From $(j,0)$, with probability $1/2$ the particle at
site $j$ is chosen, but it is unable to move due to the exclusion interaction
and the presence of the other particle at the hub $0$, which is the only site
available. If the particle at the hub is chosen, then it moves to site $i
\neq j$ with probability $1/L$.

The detailed balance condition
\begin{equation}
p_{(0,i)}  P\left[ (0,i) \to (j,i) \right] = p_{(j,i)}  P\left[ (j,i) \to (0,i)
\right]
\end{equation}
then shows that $p_{(0,i)} = L p$.
Since the normalisation condition $\sum_{\s} \prob{\s} = 1$ must be satisfied,
we
have
\begin{equation}
 2 L^2 p + L(L-1) p = 1,
\end{equation}
and hence $p = 1 / (3L^2 - L)$.

We have thus found the equilibrium probability $\prob{\s}$ of each configuration
$\s$.
To find the mean occupation number $\mean{n_i}$ of site $i$, we must sum over
configurations:
\begin{equation}
 \mean{n_i} = \sum_{\s} n_i(\s) \prob{\s},
\end{equation}
where $n_i(\s)$ is the occupation number of site $i$ in the configuration $\s$.
In the case of exclusion, $n_i$ can only take the values $0$ and $1$, so that
the mean occupation of the hub is given by a sum over those configurations $\s$
which have a particle in the hub, giving
\begin{equation}
 \mean{n_0} = 2 L^2 p.
\end{equation}
The mean occupation of a site $i \neq 0$ is similarly given by a weighted sum
over those configurations which have a particle in site $i$:
\begin{align}
 \mean{n_i} &= p_{(i,0)} + p_{(0,i)} + \sum_{j=1,\ldots,L; j \neq i}
\left[ p_{(i,j)} + p_{(j,i)} \right] \\
&= 2Lp + 2(L-1) p = (4L - 2)p.
\end{align}
The normalisation $\mean{n_0} + \sum_{i=1}^L \mean{n_i} = 2$ is then correctly
satisfied, and
we have also checked these exact results with numerical simulations (not shown).

In such structured networks, we can also proceed to obtain results for more
detailed features of the probability distributions, such as higher moments.

\subsubsection{Mean encounter time}

Identifying particle $1$ as the distinguished walker,
the probability $\prob{E}$ of its encounter set
may be calculated in a similar way to that in
sec.~\ref{sec:reg-lattice-exclusion-mean-encounter}, as
the sum over all configurations such that $1$ has a neighbour, weighted by the
probability that an encounter occurs, i.e., that $1$ interacts with the
neighbour. In this simple system, encounters can occur only with configurations
of the form $(0,i)$ or $(i,0)$, for which one
of the particles is at the hub. In this case, the probability that
the two particles interact is $1/2 + 1/(2L)$, since the particle not at the hub
always tries to jump towards the hub, whereas the particle at the hub
usually jumps towards an empty node.
\begin{align}
 \prob{E} &=  \sum_{i} \left[p_{(i,0)} + p_{(0,i)} \right]  \left[\frac{1}{2} +
\frac{1}{2L} \right] \\
 &= 
p (L^2 + L), 
\end{align}
giving
\begin{equation}
 \mean{\tau} = \frac{1}{p(L^2 + L)} 
 = 3 - \frac{4}{L+1}.
\end{equation}

For a network with exclusion, the total number of spatial configurations is $V!
/ (V-N)!$, so
that for arbitrary 
networks this kind of calculation becomes intractable. Nonetheless, for networks
which are small and/or have enough structure it can be carried out relatively
easily.

\subsection{Equilibrium distribution in the large-system approximation}

For systems with many nodes, for which the above direct method is
impractical, it is instead necessary to turn to
an approximation in which we 
assume that the occupation numbers of neighbouring sites are
\emph{independent} 
\cite{ArgyrakisBenAvrahamPriorityDiffnComplexNetworksPRE2008,
DeMouraFermiDiracStatsComplexNetworksPRE2005}. This is valid
when the system is large, or in a grand-canonical
situation, where the number of particles in the system can fluctuate about a
mean value, 
since in a finite system with a fixed number of particles, the presence or
absence
of a particle at a site $i$ affects the conditional probability to have
a particle at site $j \neq i$, given the occupation number of site $i$.

To derive the equilibrium distribution in this
approximation, the method of
ref.~\cite{ArgyrakisBenAvrahamPriorityDiffnComplexNetworksPRE2008} can be
applied. As shown below, the results
which were first obtained in
ref.~\cite{DeMouraFermiDiracStatsComplexNetworksPRE2005} are recovered, in a
more direct way.

\subsubsection{Neighbouring sites}
Let $p_i$ be the probability that site $i$ is occupied in equilibrium. Due to
the exclusion interaction, the occupation number $n_i$ of site $i$ is either
$0$ or $1$, so that its mean is $\mean{n_i} = p_i$.
Thus $\sum_i p_i = N$, the total number of particles present in the system.

Suppose that the particle on site $i$ is chosen to move towards a
neighbouring site $j$. The probability that the trial site is unoccupied is
$(1-p_j)$ under the independence approximation, and the probability that the
direction towards site $j$ is chosen is
$1/k_i$. We thus obtain 
$P_{i \to j} = \frac{1}{N k_i} (1-p_j)$.

Here the assumption of independence of occupation states has already been used.

In equilibrium, the detailed balance condition
 $p_i P_{i \to j} = p_j P_{j \to i}$
gives
\begin{equation}
 \frac {1}{k_i} p_i (1-p_j)  = \frac {1}{k_j} p_j (1-p_i).
\end{equation}
Rearranging to collect all terms in $p_i$ and $p_j$ on opposite sides of the
equation we see that
\begin{equation}
 \frac{p_i}{k_i(1 - p_i)} = \frac{p_j}{k_j(1 - p_j)},
\end{equation}
and hence
\begin{equation}
 \frac{p_i}{k_i(1 - p_i)} = C,
\end{equation}
where $C$ is a site-independent constant.
Finally we obtain
\begin{equation} 
 p_i = \frac{C k_i}{1 + C k_i} = \frac{1}{1 + A k_i^{-1}},
 \label{eq:A}
\end{equation}
where $A$ is another constant, as was found in
ref.~\cite{DeMouraFermiDiracStatsComplexNetworksPRE2005}.
The constant $A$ is determined by the normalisation condition $\sum p_i = N$,
and thus depends on the entire set of degrees $\{k_i\}$.

\subsubsection{Single site}

A single-site variant of the above Markov chain method gives an alternative
derivation.
Consider a site $i$ with degree $k_i$.
Let $p_0^{(i)} = 1 - p_i$ and  $p_1^{(i)} = p_i$ be the equilibrium
probability that the site is empty or
occupied, respectively.
We first need a 
mean-field type estimate of the  transition probabilities from empty to
occupied, $P_{0 \to 1}^{(i)}$, and vice versa, $P_{1 \to 0}^{(i)}$.

Due to the way the network is constructed, following a given link from a given
node leads to a new node $j$ with probability $k_j / K$ which depends on $j$,
since node $j$ has $k_j$ incoming edges
\cite{DeMouraFermiDiracStatsComplexNetworksPRE2005}. The particle at node $j$
then jumps to site $i$ with probability $1/k_j$, giving
\cite{ArgyrakisBenAvrahamPriorityDiffnComplexNetworksPRE2008}
\begin{align}
P_{0 \to 1}^{(i)} &= \frac{1}{N} \sum_j \frac{k_j}{K} p_j \frac{1}{k_j}
=\frac{1}{N} k_i \frac{\rho}{\mean{k}},
\end{align}
The second equality follows since $\sum_j p_j = N$ and $N/K = \rho/\mean{k}$.

We calculate $P_{1 \to 0}^{(i)}$ by arguing similarly. If the particle at site
$i$ is selected, with probability $1/N$, then it can attempt to move to any of
the $k_i$ neighbours, each with probability $1/k_i$. 
The neighbour is
site $j$ with probability $k_j / K$.
The move is successful only if the neighbour is empty, due to the
exclusion, and occurs with probability $1-p_j$, giving
\begin{equation}
 P_{1 \to 0}^{(i)} = \frac{1}{N} \sum_j \frac{k_j}{K} (1-p_j).
\end{equation}
Note the extra factor $(1-p_j)$ compared
to the expression  for independent
dynamics in ref.~\cite{ArgyrakisBenAvrahamPriorityDiffnComplexNetworksPRE2008}.

Detailed balance gives $(1-p_i) P_{0 \to 1}^{(i)} = p_i P_{1 \to
0}^{(i)}$, from which
we finally obtain
\eqref{eq:A} again, but now with an expression for $A$:
\begin{equation}
 A = \frac{\mean{k}}{\rho} \left( 1 - \frac{\sum_j k_j p_j}{K} \right) =
\frac{1}{N} \sum_j k_j (1-p_j).
\label{eq:expression-for-A}
\end{equation}
Although this equation appears to give new information, 
in fact it turns out to be equivalent to the 
normalisation condition.

Unfortunately, it does not seem to be possible to solve this equation exactly to
find $A$ and the $p_i$ explicitly.  In
ref.~\cite{DeMouraFermiDiracStatsComplexNetworksPRE2005}, $A$ was found
numerically by solving the normalisation equation $\sum_j p_j = N$. 
This gives no insight into the quantity $A$, however.

An alternative is to find approximations to $A$.
A first approximation is obtained by taking all $k_i$ equal to $\mean{k}$ in
\eqref{eq:expression-for-A},
giving
\begin{equation}
 A^{(0)} = \frac{1-\rho}{\rho} \mean{k}. 
 \label{eq:A0}
\end{equation}
As shown below in fig.~\ref{fig:visits}, this already gives a reasonable
approximation to the
distribution $p_i$ for networks for which the deviation of the $k_i$ from their
mean is small, although the corresponding $p_i^{(0)}$ calculated using this
value for $A$ do not satisfy the
normalisation condition.

Further approximations may be obtained -- either analytically or numerically --
by an iterative scheme based on \eqref{eq:expression-for-A} and with the initial
value \eqref{eq:A0} for $A$:
\begin{align}
 A^{(n+1)} &\defeq \frac{1}{N} \sum_j k_j (1 - p_j^{(n)});\\
   p_i^{(n)} &\defeq \frac{1}{1 + A^{(n)} k_i^{-1}},
\end{align}
giving
\begin{equation}
  A^{(n+1)} =\frac{1}{N}\sum_j \frac{1}{[A^{(n)}]^{-1} + k_j^{-1}}.
\end{equation}
This iteration, which is easily implemented computationally, quickly converges
to a fixed point which gives the numerically exact value of $A$ and of the
$p_j$ for the given
degree sequence, and thus provides an alternative numerical method to that used
in
ref.~\cite{DeMouraFermiDiracStatsComplexNetworksPRE2005}.

\subsection{Mean encounter time in the large-system approximation}

The calculation of the mean encounter time in the
large-system approximation, supposing that the occupation
probabilities of neighbouring sites are independent of each other, proceeds as
follows..

For the distinguished walker $0$ to have an encounter, it first must be at some
site $i$, which occurs with probability $p_i / N$ (giving a total 
probability
$1$ to be at some site).
There are two possibilities for such encounters: either
walker $0$ is
chosen to jump, in which case it has an encounter
if the trial site is occupied,  or another
walker attempts to jump onto the site occupied by walker $0$.

The probabilities for these two possibilities are calculated in the
same way as the transition probabilities in the previous section.
Again denoting by $E$ the encounter set of the distinguished walker, we obtain
\begin{align}
  \prob{E} = \sum_i \frac{p_i}{N} \left[ \sum_j \frac{k_j}{K} p_j +
k_i \sum_j
\frac{k_j}{K} p_j \frac{1}{k_j} \right]
= \frac{2}{K} \sum_j k_j p_j,
\end{align}
and hence
\begin{equation}
 \mean{\tau} = \frac{V\mean{k}}{2 \sum_j k_j p_j}.
\end{equation}
We remark that this calculation is basically that of 
the \emph{jamming
probability} studied in
ref.~\cite{DeMouraFermiDiracStatsComplexNetworksPRE2005}, i.e., the
probability
that the particle which attempts to move is jammed (blocked)
\cite{DeMouraFermiDiracStatsComplexNetworksPRE2005} -- that is, that an
encounter occurs.

\section{Numerical results}

\label{sec:numerical}

In this section, the analytical results obtained in
previous sections are compared with the
results of numerical simulations on two different types of network, one regular
and one complex.

\subsection{Regular network: linear chain}
First consider 
a regular network, consisting of 
a linear chain of $V=100$ sites, where each site is connected to its two nearest
neighbours, with periodic boundary conditions.

\begin{figure}
\includegraphics{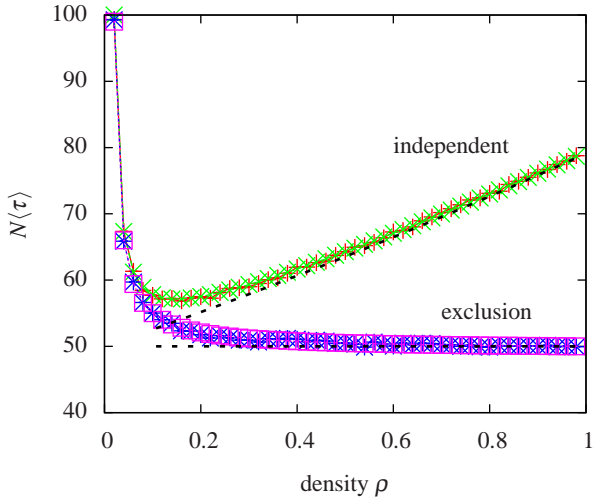} 
\caption{Mean encounter time $N\mean{\tau}$ on a 
linear chain of
length
$V=100$, as a function of the number of particles $N$, with and without
exclusion. Numerical
results, evaluated as a mean over $10^8$ steps, are
compared to the analytical results;
lines are shown as a guide for the eye.  Error bars are of the order of the
symbol size.  There is excellent agreement in both cases.
Dashed lines show the asymptotic behaviour.
}
 \label{fig:chain}
\end{figure}

The mean encounter time of a distinguished walker for dynamics
both with and without exclusion on this chain are shown in fig.~\ref{fig:chain}
as a function of the total number of walkers, $N$, between
$2$ and $V$.  To distinguish the two cases, the time is shown as $N
\mean{\tau}$, i.e., as a raw number of steps, rather than as a number of sweeps.
The analytical and numerical results in both cases agree very
well.

The figure shows that the mean encounter time (in sweeps) depends very
little on the dynamics. The mean encounter time in the case of exclusion
dynamics is generally slightly
shorter, which we can attribute to the fact that the particles must be spread
out more uniformly through the system in this case due to the exclusion
interaction.

Note that at first glance the Kac result
\eqref{eq:exact-encounter-time-reg-exclusion} does not hold for a
one-dimensional 
dynamics with strict exclusion, since this result
assumes ergodicity, i.e., that it any
configuration can be reached from any other, which is not the case
due to the one-dimensional nature of the system: each
walker is always confined between the same two neighbours. However, the result
is in fact valid. 
This is because the mean
encounter
time is a single-particle quantity, which can be calculated by averaging
over all particles in the system.  The result for the \emph{global} encounter
time (taking into account encounters of any particle) will be the same in the
ergodic and non-ergodic cases, since each time between two encounters is
unaffected, but may be assigned to a different walker. This then implies
equality also for the encounter times of a distinguished walker.

\subsection{Complex networks: random graphs with power-law degree
distributions}

\begin{figure}
\includegraphics{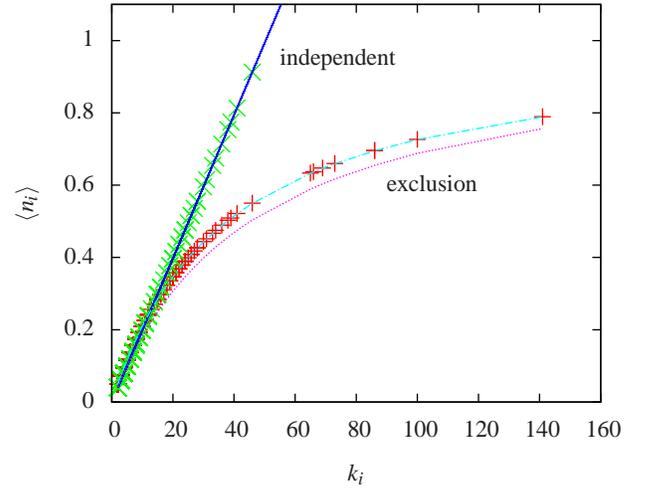} 
\caption{Mean occupation number $\mean{n_i}$ for each site $i$, as a function
of the degree $k_i$ of the site, for dynamics with and without exclusion, on a
single random network with power-law degree distribution $P(k) \sim
k^{-\alpha}$ with $\alpha=2.5$. The network has
$V=1000$ nodes and mean degree $\mean{k}=5.036$.
The numerical data for each site is shown as a symbol, 
and the curves show the analytical results; the lowest curve is the
zeroth-order approximation $p_i{(0)}$ in the exclusion case.
The complete curve in the case of independent dynamics continues to grow
linearly for larger $k_i$ (not shown).
}
 \label{fig:visits}
\end{figure}

\begin{figure*}
\subfigure[]{
\includegraphics{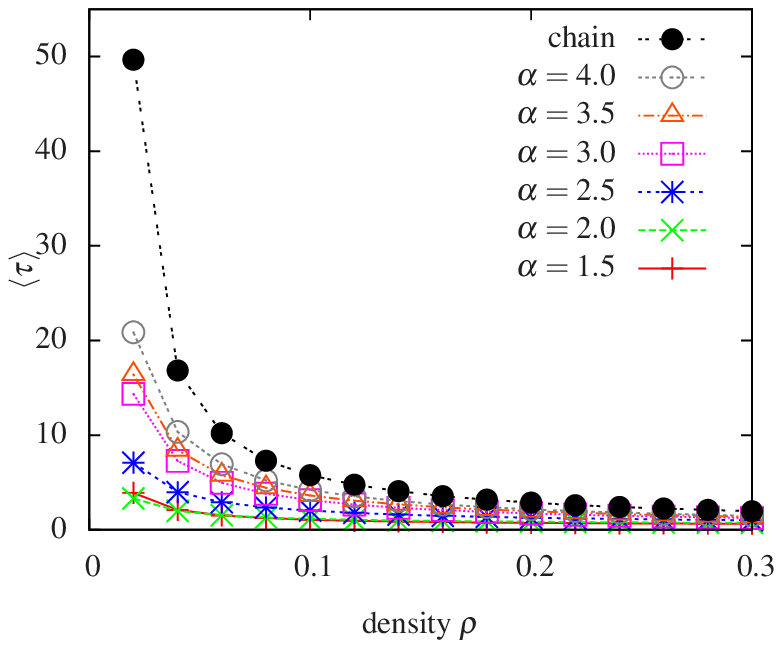} 
}
\subfigure[]{
\includegraphics{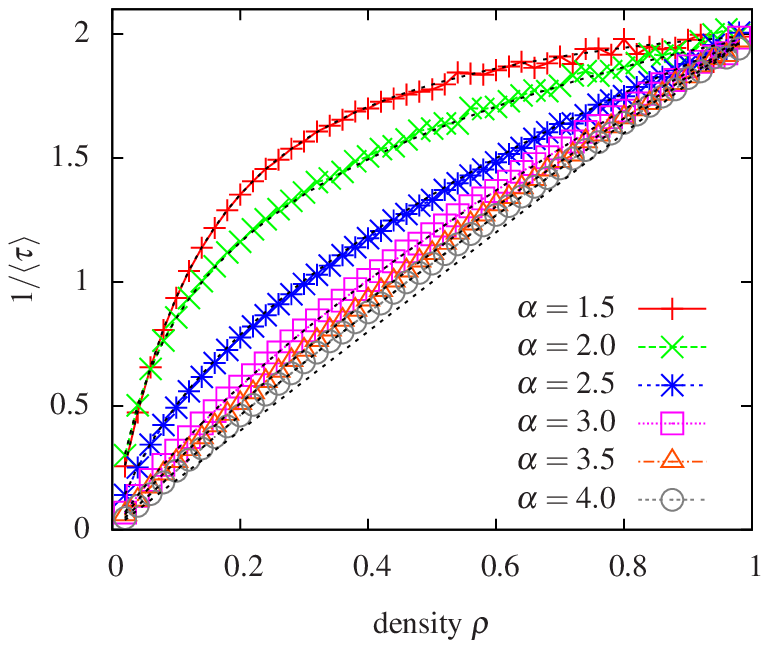} 
}
\caption{(a) Mean encounter time $\mean{\tau}$ of a distinguished walker on
networks with power-law degree distribution $P(k) \sim k^{-\alpha}$, for
different values of $\alpha$ with $V=1000$ sites and for a chain with $V=100$,
as a function of the density, $\rho$. Only data for exclusion dynamics are
shown; data for independent walkers are very close to these.
To highlight the differences between the curves, (b) shows $1/\mean{\tau}$ 
for different $\alpha$ compared to the analytical results,
drawn with black dotted lines.
The lowest dashed line shows the mean-field result for comparison.
}
 \label{fig:power-law-times}
\end{figure*}

The second case is that of 
random networks with a power-law degree distribution
$P(k) \sim k^{-\alpha}$. These are  
generated according to the prescription in
ref.~\cite{NewmanStrogatzWattsRandomGraphsArbDegreeDistnPRE2001}:
(i) a degree sequence $(k_i)_{i=1}^V$ is generated from the distribution,
rejecting each $k_i$ if it does not satisfy $2 \le k_i \le N$; (ii) 
$k_i$ ``stubs''
are generated at each node $i$; and (iii)
pairs of stubs are chosen at random to be connected.
This method 
gives networks which in general include 
self-links from a given node back to itself, as
well as 
multiple links between nodes
\cite{BogunaGenerationUncorrelatedRandomScaleFreeNetworksPRE2005}. 
Since both
the random-walk dynamics
and our analytical results take these into account, no attempt was made to
remove
them from the network, as is done in
ref.~\cite{BogunaGenerationUncorrelatedRandomScaleFreeNetworksPRE2005} for
example; rather, this gives a more stringent test of the analytical
results. 
The
imposed minimum degree of $2$ at each node ensures that the resulting network
is connected with probability one
\cite{BogunaGenerationUncorrelatedRandomScaleFreeNetworksPRE2005}. 
 
 Power-law networks with
smaller values of $\alpha$ have more nodes of 
high degree, and in particular a few very highly-connected hubs.
Particles will concentrate at or near these hubs, and so intuitively
this will lead to shorter mean encounter times.
For an infinite system, the degree distribution has a well-defined mean if and
only if 
$\alpha > 2$, but for a finite network we can also consider $\alpha < 2$. We do,
however, impose the total number of
sites as a cutoff for the maximum allowed degree.

Figure~\ref{fig:visits} shows a comparison of numerical and analytical 
results for the mean occupation number $\mean{n_i}$ (which is
equal
to $p_i$ in the case of exclusion dynamics) in
the case $\alpha = 2.5$, for
dynamics with and without exclusion. We see
that the zeroth-order approximation $p_i{(0)}$ already provides a good
approximation for exclusion dynamics, even though the values of
$k_i$ cover a wide range of values, including far from the mean $\mean{k}$. 
The converged $p_i$ agree very well indeed with the numerical values, as was
already found in ref.~\cite{DeMouraFermiDiracStatsComplexNetworksPRE2005}.

\begin{figure}
\includegraphics{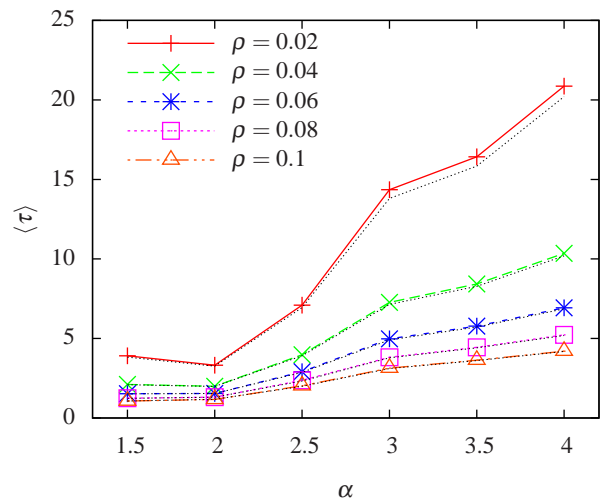} 
\caption{Mean encounter time $\mean{\tau}$ as a function of the
power-law exponent $\alpha$, for different densities $\rho$, on networks of
size $V=1000$.  The black dotted lines show the analytical results for
comparison.
}
 \label{fig:times-fn-alpha}
\end{figure}

Figure~\ref{fig:power-law-times} shows the mean encounter times for 
networks with different power-law degree distributions. Part (a) shows
$\mean{\tau}$, and part (b) shows $1/\mean{\tau}$ to
exhibit more clearly the differences between networks with different $\alpha$.
The main observation is that networks with smaller $\alpha$, i.e., with
highly-connected hubs, indeed have lower mean encounter times. This is
highlighted in \ref{fig:times-fn-alpha}, where the encounter time
is plotted for different values of $\rho$ as a function of $\alpha$.
We also see that the exact and numerical results again agree very well.
Results for non-exclusion dynamics on the same graphs are very similar, although
slightly larger, for the same reason as in regular networks, and are not shown.

\section{Conclusions}
\label{sec:conclusions}

In conclusion, this paper has shown that it is often possible to calculate
analytically a key
quantity in systems consisting of many interacting random walkers, namely the
mean encounter time of a given particle.  This was carried out for the case of
independent walkers and for walkers with exclusion on regular and complex
networks, and the results were successfully compared to numerical simulations.

For a given graph, the mean encounter time is very similar whether dynamics
with or without exclusion is used, even though the mean occupation numbers can
be quite different. This could change with a different choice of 
interaction rule in the case of independent walkers.

At first glance, it seems that the results require averages over a very long
time to be valid, namely the time required for a given walker to explore the
whole system. In the case of a high-density system with exclusion, for example,
this timescale could be very long. In fact, however, the results are unaffected
by considering interacting particles which exchange positions, so that the
timescale required is more like that for a single particle to explore the
system when no others are present. 

The method employed can be extended to other mean
encounter
times of interest. For example, interaction times between two distinguished
walkers can be found.
The extension of these results to higher moments and the full
probability distribution of encounter times, and the effect of different network
structures on those results, are subjects for future study.

The author thanks G.~Naumis for suggesting to study the Bonabeau
model, M.~Aldana for a helpful
conversation, and H.~Larralde for useful discussions and for a reading
the manuscript critically. Financial support from the PROFIP program
of
DGAPA-UNAM is acknowledged.

\vspace*{20pt}
\appendix
\section{Intuitive derivation of the Kac recurrence theorem}

Rigorous derivations of the Kac recurrence theorem, such as can be found in
refs.~\cite{KacRecurrenceThmBullAMS1947,
KacProbabilityPhysicalSciencesBook,
CondaminRandomWalksFirstPassageConfinedPRE2007}, do not always provide 
intuition
about why the result should be true.  Here, a simple, non-rigorous
argument is given which captures the essence of the result.

To find the mean recurrence time $\mean{\taurec{A}}$ to a set $A$ in a
discrete-time, ergodic system, 
consider a long trajectory of the system, of length $T$ time steps.  If at time
$t$ the system is in $A$, then write a $1$; if it is outside $A$, then
write a
$0$, thus coding  the trajectory as a symbol sequence of $0$s
and $1$s.

At long times, $T \to \infty$, 
the proportion of $1$s in the sequence converges 
 to the equilibrium probability $\prob{A}$
that the system is inside $A$. This is the crucial part
of the argument. From a physical point of view, it is
a weak version of the Boltzmann ergodic hypothesis, but in the case of
discrete-time
stochastic processes it is made rigorous by the Kac recurrence theorem
\cite{KacRecurrenceThmBullAMS1947}.
The number of $1$s
occurring in time $T$ is thus roughly $T \prob{A}$.
Similarly, the total time spent \emph{outside} $A$ is approximately $T
\left[1-\prob{A}\right]$. 

Now consider
rearranging the list of $1$s and $0$s so that approximately the same number of
$0$s occurs between each pair of consecutive $1$s.  The mean recurrence time is
then this number of $0$s, plus $1$ for the extra step to return to
 the next $1$, giving 
\begin{equation}
\mean{\taurec{A}} = 1 + \frac{T \left[1-\prob{A}\right]}{T \, \prob{A}}
= \frac{1}{\prob{A}},
\end{equation}
which is the Kac result.

\end{document}